\documentclass[twocolumn,letter]{jpsj3}

\usepackage{color}
\usepackage{graphicx}
\usepackage{dcolumn}
\usepackage{bm}

\title{Pressure-Induced Superconductivity in Mineral Calaverite AuTe$_2$}

\author{Shunsaku~Kitagawa$^1$\thanks{E-mail address: shunsaku@crystal.kobe-u.ac.jp}, Hisashi~Kotegawa$^1$, Hideki~Tou$^1$, Hiroyuki~Ishii$^2$, Kazutaka~Kudo$^2$, Minoru~Nohara$^2$, and Hisatomo~Harima$^1$}
\inst{$^1$Department of Physics, Kobe University, Kobe 657-8501, Japan \\
$^2$Department of Physics, Okayama University, Okayama 700-8530, Japan}

\abst{The pressure dependences of resistivity and ac susceptibility have been measured in the mineral calaverite AuTe$_2$.
Resistivity clearly shows a first-order phase transition into a high-pressure phase, consistent with the results of a previous structural analysis.
We found zero resistivity and a diamagnetic shielding signal at low temperatures in the high-pressure phase, which clearly indicates the appearance of superconductivity.
Our experimental results suggest that bulk superconductivity appears only in the high-pressure phase.
For AuTe$_2$, the highest superconducting transition temperature under pressure is $T_{\rm c}$ = 2.3~K at 2.34 GPa; it was $T_{\rm c}$ = 4.0~K for Pt-doped (Au$_{0.65}$Pt$_{0.35}$)Te$_2$.
The difference in $T_{\rm c}$ between the two systems is discussed on the basis of the results obtained using the band calculations and McMillan's formula.
}

\kword{AuTe$_2$, superconductivity, structural phase transition, pressure}

\begin{document}
\maketitle

Crystal structure plays a crucial role in the appearance of superconductivity.
In iron pnictides, structural phase transition from a high-temperature tetragonal phase to a low-temperature orthorhombic phase is suppressed by chemical substitution or pressure, and superconductivity mainly appears in the tetragonal phase\cite{Y.Kamihara_JACS_2008,K.Ishida_JPSJ_2009,J.Paglione_Naturephys_2010}.
IrTe$_2$ also shows superconductivity when its structural phase transition from a high-temperature trigonal structure to a low-temperature monoclinic structure is suppressed by the substitution of Pt for Ir, Cu-intercalation, and so on\cite{S.Pyon_JPSJ_2012,K.Kudo_JPSJ_2013,J.J.Yang_PRL_2012,M.Kamitani_PRB_2013}.
Since this structural phase transition is characterized by the formation of Ir-Ir bonds along the $b$-axis, this superconductivity can be interpreted by bond-breaking-induced superconductivity.
Moreover, it has recently been found that the making and breaking of dimers or bonds change the electronic state markedly and induce superconductivity or magnetic order in several compounds\cite{D.Hirai_PRB_2012,M.Danura_JPSJ_2011,S.Jia_PRB_2010,S.Jia_NaturePhys_2011,S.Kasahara_PRB_2011}.
Therefore, it is important in the search for a new phase to investigate low-temperature properties near chemical bonding instability.

\begin{figure}[!tb]
\vspace*{0pt}
\begin{center}
\includegraphics[width=6.5cm,clip]{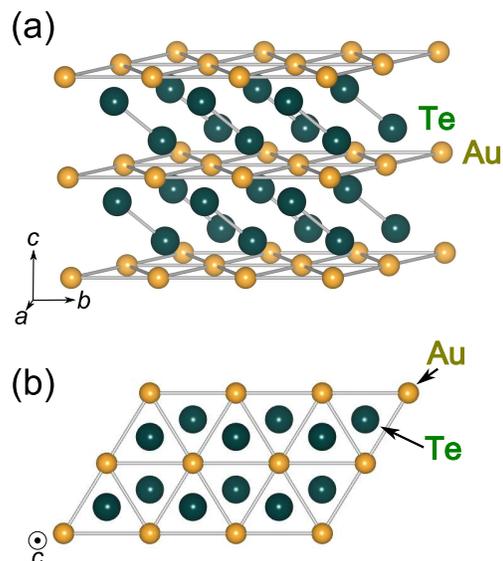}
\end{center}
\caption{(Color online)(a) Average crystal structure of monoclinic calaverite AuTe$_2$ with the space group $C2/m$.
Te$_2$ dimers are formed in the actual structure.
(b) Top ($c$-axis) view of the average structure of monoclinic AuTe$_2$.
For drawings of the crystal structure, the software Vesta was used\cite{K.Momma_JAC_2008}. }
\label{Fig.1}
\vspace*{-20pt}
\end{figure}

The mineral calaverite AuTe$_2$ has a incommensurately modulated monoclinic structure.
The average structure of AuTe$_2$ is a monoclinically distorted CdI$_2$-type structure with the space group $C2/m$ (No. 12), as shown in Fig.~\ref{Fig.1}\cite{G.Tunell_ActaCryst_1952}.
The two-dimensional Au planes and Te layers are stacked alternately.
In this average structure, Te atoms form zigzag chains with an interatomic distance of 3.20 \AA.
In the actual structure, incommensurate modulation ($\bm{q}$ = - 0.4076$\bm{a}^*$ + 0.4479 $\bm{c}^*$) induces the breaking up of zigzag chains and the formation of isolated Te$_2$ dimers with a short interatomic distance of 2.88 \AA\cite{A.Janner_ActaCryst_1989}.
Although the charge density wave state or mixed-valence state of Au$^{+}$ (5$d^{10}$) and Au$^{3+}$ (5$d^{8}$) is proposed to be the origin of such incommensurate modulation\cite{W.J.Schutte_ActaCryst_1988}, X-ray photoelectron spectroscopy and first principles calculation indicate the homogeneous monovalent Au$^{+}$ state of AuTe$_2$ to be the case\cite{A.v.Triest_JPhys_1990,B.C.H.Krutzen_JPhys_1990}.
In addition, the calculated Fermi surface of the average structure shows that the modulation cannot be understood in terms of Fermi-surface nesting\cite{B.C.H.Krutzen_JPhys_1990}.
The origin of the incommensurate modulation remains unclarified.
The incommensurately modulated structure changes to a modulation-free trigonal (distortion-free) CdI$_2$-type structure at room temperature at 2.5~GPa with pressure application\cite{K.Reithmayer_ActaCryst_1993}.
In the high-pressure trigonal phase, Te$_2$ dimers break up.
Quite recently, it has been reported that Pt substitution induces structural phase transition to a trigonal phase and that dimer-breaking-induced superconductivity appears at a superconducting transition temperature $T_{\rm c}$ = 4.0~K at ambient pressure\cite{K.Kudo_JPSJ_2013}.
Therefore, it is expected that superconductivity will also appear in the high-pressure trigonal phase.

In this study, we measured the pressure dependences of resistivity and ac susceptibility up to $\sim$ 4.2~GPa and found superconductivity at $T_{\rm c} \simeq 2$~K in the high-pressure phase.
We also performed first-principles band calculations of the average structure of the low-pressure monoclinic phase and trigonal structure.
The calculated Sommerfeld coefficient $\gamma$, which is proportional to the density of states at around the Fermi energy $D(E_{\rm F})$ in the low-pressure monoclinic phase, is clearly larger than that estimated from the specific heat measurement.
This disagreement suggests that the calculation of the average structure does not reproduce the actual band structure, indicating the importance of Te$_2$ dimers due to the incommensurate modulation of the crystal structure for the electronic state.

Single crystals of AuTe$_2$ were grown by heating a mixture of Au (99.99\%) and Te (99.99\%). 
The stoichiometric amount of the powder mixture was sealed in an evacuated quartz tube. 
The ampules were heated to 470 $^{\rm o}$C at 18 $^{\rm o}$C/h, then heated to 600 $^{\rm o}$C at 6.5 $^{\rm o}$C/h, and finally cooled to room temperature at 11.5 $^{\rm o}$C/h. 
The obtained single crystals were confirmed to be single-phase AuTe$_2$ by powder X-ray diffraction analysis using a Rigaku RINT-TTR III X-ray diffractometer with Cu K$\alpha$ radiation. 
Electrical resistivity ($\rho$) and ac susceptibility ($\chi_{\rm ac}$) measurements at high pressures of up to $\sim$ 4.2 GPa were carried out using an indenter cell\cite{T.C.Kobayashi_RSI_2007}.
Temperature dependence measurements at each pressure were carried out in the pressure-increasing process.
Electrical resistivity was measured by a standard four-probe method. 
Ac susceptibility was measured by the mutual-inductance technique using a lock-in amplifier.
The sample was shaped into a block with dimensions of approximately 1.0 $\times$ 0.4 $\times$ 0.1 mm$^3$ for electrical resistivity measurements.
For ac susceptibility measurements, we used powdered samples.
We used Daphne 7474 as the pressure-transmitting medium\cite{K.Murata_RSI_2008}. 
The applied pressure was estimated from the $T_{\rm c}$ of the lead manometer.
Since the applied pressure decreases by approximately 0.2~GPa on cooling, the actual pressure at high temperatures is larger than the displayed pressure.

\begin{figure}[!tb]
\vspace*{0pt}
\begin{center}
\includegraphics[width=8cm,clip]{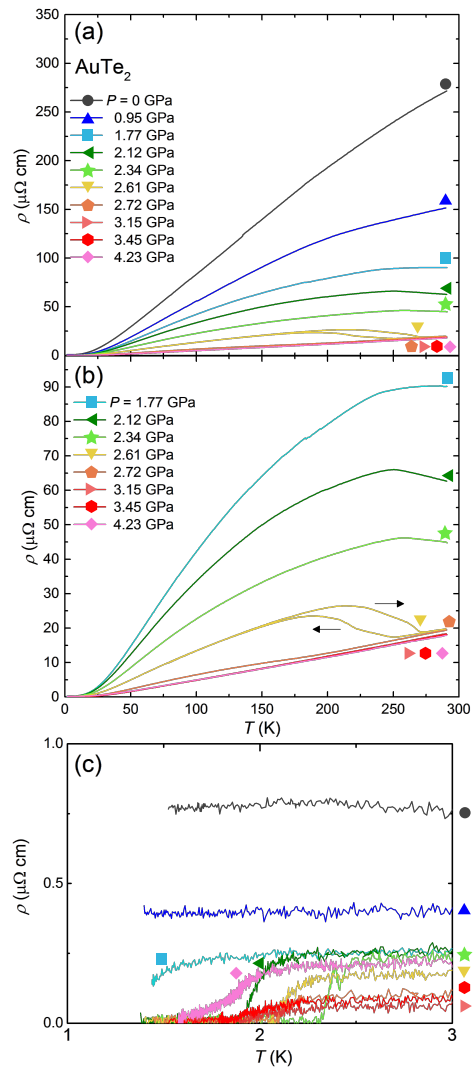}
\end{center}
\vspace*{-20pt}
\caption{(Color online)(a) Temperature dependence of resistivity at several pressures in AuTe$_2$.
In the entire pressure range, resistivity shows a metallic behavior.
By applying pressure, $\rho(T)$ in the entire temperature range is drastically suppressed with increasing pressure up to $\simeq$ 2.6~GPa.
(b) Temperature dependence of resistivity above 1.77~GPa.
(c) Temperature dependence of resistivity below 3~K. $\rho(T)$ drops and shows zero resistivity above 2.12 GPa.}
\label{Fig.2}
\vspace*{-20pt}
\end{figure}

Figure~\ref{Fig.2} shows the temperature dependence of $\rho(T)$ at several pressures.
At ambient pressure, resistivity monotonically decreases on cooling down to 1.5~K, which is consistent with a previous report\cite{K.Kudo_JPSJ_2013}.
A residual resistivity $\rho_0$ = 0.78 $\mu\Omega\cdot$cm and a residual resistivity ratio $\rho(290~{\rm K})/\rho_0 \simeq 350$ indicate good sample quality.
By applying pressure, $\rho(T)$ in the enire temperature range is drastically suppressed with increasing pressure up to $\simeq$ 2.6~GPa, as shown in Fig.~\ref{Fig.2} (a).
$\rho(290~{\rm K})$ at 2.61~GPa becomes approximately 14 times smaller than that at ambient pressure.
In a normal metal, $\rho$ can be described as $\rho = m^*/(e^2 n \tau )$, where $m^*$ is the effective mass of an electron, $e$ is the elementary charge, $n$ is the charge carrier density, and $\tau$ is the electron relaxation time.
Therefore, the decrease in $\rho$ originates from the increase in $n$ and/or the decrease in the electron scattering rate 1/$\tau$, indicative of the marked change in the electronic state with pressure.
Above $\sim$~3~GPa, $\rho(T)$ does not change so much with pressure.

\begin{figure}[!tb]
\vspace*{0pt}
\begin{center}
\includegraphics[width=8cm,clip]{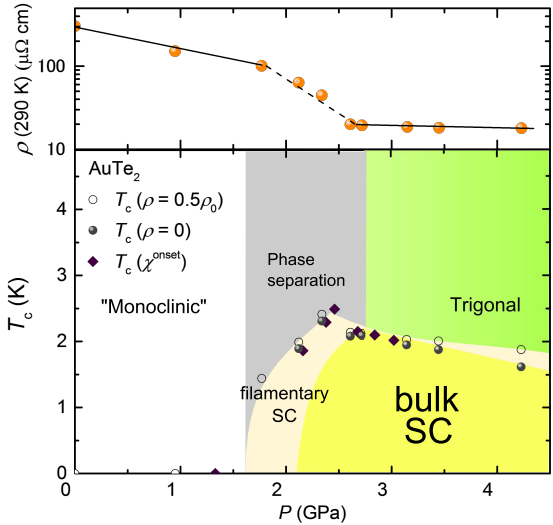}
\end{center}
\caption{(Color online)(Upper panel) Pressure dependence of $\rho(290~{\rm K})$.
$\rho(290~{\rm K})$ is drastically suppressed toward 2.6~GPa.
Since the applied pressure decreases by approximately 0.2~GPa on cooling, the actual pressure at 290~K is larger than the displayed pressure.
(lower panel) $T-P$ phase diagram for AuTe$_2$. Open(filled) circles represent $T_{\rm c}$ determined by $\rho = 0.5\rho_0 (0)$ and filled diamonds indicate $T_{\rm c}$ determined by the onset of the diamagnetic shielding signal.
The crystal symmetry inferred from a previous report\cite{K.Reithmayer_ActaCryst_1993} is shown.
}
\label{Fig.4}
\vspace*{0pt}
\end{figure}

\begin{figure}[!tb]
\vspace*{0pt}
\begin{center}
\includegraphics[width=8.5cm,clip]{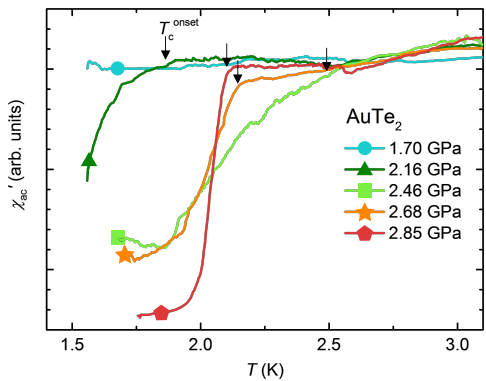}
\end{center}
\caption{(Color online) Temperature dependence of the real part of $\chi_{\rm ac}$ at several pressures.
The diamagnetic shielding fraction is small and the transition width is broad below 2.7~GPa, suggesting that phase separation of the superconducting region and the non-superconducting region exists.
The arrows indicate the onset temperature of the superconducting transition.}
\label{Fig.3}
\vspace*{-5pt}
\end{figure}

In the pressure range between 1.77 and 2.72~GPa, $\rho(T)$ shows a broad peak structure at high temperatures.
This behavior becomes noticeable at 2.61 GPa.
At 2.61 GPa, $\rho(T)$ markedly changes at $\simeq$ 250~K on cooling and at $\simeq$ 270~K on warming, as shown in Fig.~\ref{Fig.2}(b), similarly to the case of IrTe$_2$\cite{S.Pyon_JPSJ_2012}.
This first-order phase transition might originate from structural phase transition.
The pressure dependence of $\rho(290~{\rm K})$ also suggests the occurrence of the structural phase transition at $\simeq$ 2.6~GPa, as shown in the upper panel of Fig.~\ref{Fig.4}, which is consistent with a previous report\cite{K.Reithmayer_ActaCryst_1993}.
The anomaly in $\rho$ becomes smaller and the temperature for the anomaly decreases at 2.74~GPa.

At low temperatures, zero resistivity is observed above 2.12 GPa.
$T_{\rm c}$ is determined to be the temperature at which zero resistivity is obtained.
With increasing pressure, $T_{\rm c}$ increases up to 2.34 GPa and decreases above it.
The maximum of $T_{\rm c}$ is 2.3~K at 2.34 GPa.
Below 2.7~GPa, the superconducting transition is broad and the diamagnetic shielding fraction is smaller than that in the high-pressure phase, as shown in Fig.~\ref{Fig.3}.
It seems that a phase separation of the low- and high-pressure phases exists between $\sim$ 1.7 and $\sim$ 2.7~GPa and that superconductivity appears only in the high-pressure phase.
Actually, an anomaly in $\rho(T)$ corresponding to the structural phase transition is observed even below 2.6~GPa, as shown in Fig.~\ref{Fig.2}(b).

\begin{figure}[!tb]
\vspace*{5pt}
\begin{center}
\includegraphics[width=7cm,clip]{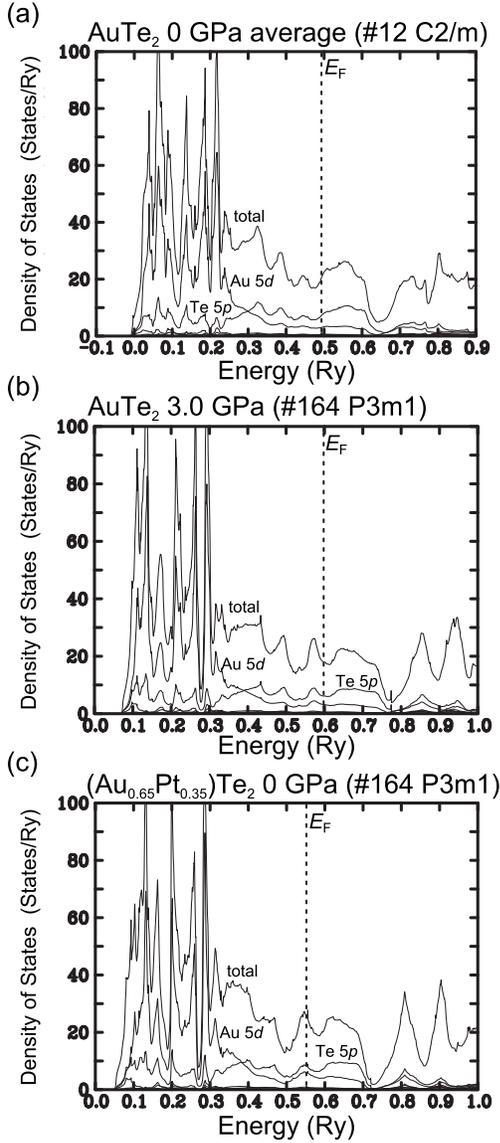}
\end{center}
\caption{Partial density of states of low-pressure monoclinic phase (a), high-pressure trigonal phase (b), and 35\%-Pt-doped trigonal phase (c) of AuTe$_2$.
In the low pressure phase, we calculate the electronic state of the average structure.
The dotted lines represent the Fermi energy.
We put atoms of $Z$ = 78.65 at Au sites for the calculation of (Au$_{0.65}$Pt$_{0.35}$)Te$_2$ (virtual crystal approximation).
Here, $Z$ is the atomic number. 
The disagreement between the calculated $\gamma$ and experimental $\gamma$ suggests that the calculation of the average structure does not reproduce the actual band structure.}
\label{Fig.5}
\end{figure}

Figure~\ref{Fig.4} shows the $T-P$ phase diagram in AuTe$_2$. 
$T_{\rm c}$ is determined by resistivity and ac susceptibility measurements.
The pressure dependence of $\rho(290~{\rm K})$ is also plotted in the upper panel of Fig.~\ref{Fig.4}.
The resistivity slope changes at $\simeq$ 1.7 and $\simeq$ 2.6~GPa.
As mentioned above, there is a phase separation region between the low- and high-pressure phases.
This is consistent with the $x$ dependence of the lattice parameters of the Pt-doped system (Au$_{1-x}$Pt$_x$)Te$_2$\cite{K.Kudo_JPSJ_2013}.
We estimated a phase separation region from the resistivity slope.
$T_{\rm c}$ in the high-pressure phase monotonically decreases with increasing pressure.


In order to discuss the variation in the electronic structure by structural phase transition, we performed first-principles band-structure calculations in the low-pressure monoclinic phase and high-pressure trigonal phase of AuTe$_2$, on the basis of local density approximation and the FLAPW method.
Figure~\ref{Fig.5} shows the energy dependences of the partial density of states in the low-pressure monoclinic phase and high-pressure trigonal phase.
In the low-pressure phase, we calculate the electronic state of the average structure.
For comparison, the partial density of states of (Au$_{0.65}$Pt$_{0.35}$)Te$_2$ (trigonal structure) is also displayed.
We put atoms of $Z$ = 78.65 at Au sites for the calculation of (Au$_{0.65}$Pt$_{0.35}$)Te$_2$ (virtual crystal approximation).
Here, $Z$ is the atomic number.
Structural parameters taken from the literature are used\cite{K.Reithmayer_ActaCryst_1993,K.Kudo_JPSJ_2013}.
The calculated $\gamma$ in the low-pressure monoclinic phase [$\gamma_{\rm cal}$ = 3.6 mJ/(mol$\cdot$ K$^2$)] is approximately three times larger than that estimated from the specific heat measurement [$\gamma_{\rm exp}$ = 1.1 mJ/(mol$\cdot$ K$^2$)]\cite{K.Kudo_JPSJ_2013}.
Generally, the experimental $\gamma$ becomes larger than the calculated $\gamma$ owing to electron-phonon and/or electron-electron interaction.
Therefore, this disagreement suggests a largely different electronic structure between average and modulated structures.
The calculation of the average structure does not reproduce the actual band structure.
Since the $\gamma_{\rm cal}$ in the lowest-order commensurate approximate super lattice with $\bm{q} = -\frac{1}{2} \bm{a^{*}} + \frac{1}{2} \bm{c^{*}}$ is almost the same as that in the average structure of the low-pressure phase\cite{B.C.H.Krutzen_JPhys_1990},
the lowest-order commensurate approximate super lattice is not sufficient for the approximation of the actual structure.
Isolated Te$_2$ dimers are formed by the incommensurability of structural modulation and Te$_2$ dimers affect the electronic structure.

\vspace{1cm}
\begin{table}[b]
\caption[]{$\gamma_{\rm cal}$ and $\gamma_{\rm exp}$ values. 
The $\gamma_{\rm cal}$ at the ambient pressure of AuTe$_2$ is the value for the average structure.
- indicates that the data does not exist.} 
\label{table:1}
\vspace{1cm}
\begin{tabular}{rcc}
\hline
                   & $\gamma_{\rm cal}\left(\frac{\rm mJ}{{\rm mol}\cdot {\rm K}^2}\right)$& $\gamma_{\rm exp}\left(\frac{\rm mJ}{{\rm mol}\cdot {\rm K}^2}\right)$ \vspace{0.1cm}\\ \hline
AuTe$_2$ at 0 GPa   & 3.6 & 1.1 \\
AuTe$_2$ at 3.0 GPa & 3.1 & - \\
(Au$_{0.65}$Pt$_{0.35}$)Te$_2$ & 4.5 & 5.5 \\
\hline
\end{tabular}
\label{Tab.1}
\end{table}

\begin{figure}[!tb]
\vspace*{0pt}
\begin{center}
\includegraphics[width=8cm,clip]{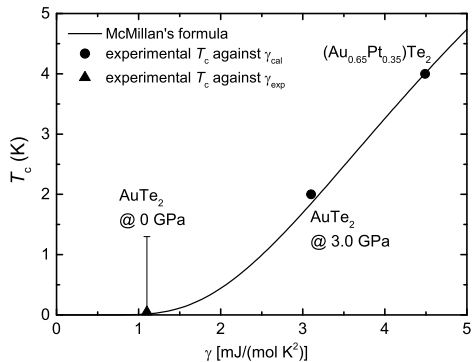}
\end{center}
\caption{Sommerfeld coefficient $\gamma$ dependence of $T_{\rm c}$ in AuTe$_2$.
The solid line indicates $T_{\rm c}$ calculated from McMillan's formula.
$\lambda(\gamma)$ and $\mu^{*}(\gamma)$ are normalized by the value in (Au$_{0.65}$Pt$_{0.35}$)Te$_2$.
The circles mean experimental $T_{\rm c}$ against $\gamma_{\rm cal}$ except for AuTe$_2$ at ambient pressure.
In AuTe$_2$ at ambient pressure, $\gamma_{\rm exp}$ is used since the calculation for the average structure does not reproduce the actual band structure}.
\label{Fig.6}
\vspace*{-20pt}
\end{figure}

Finally, we argue the difference in $T_{\rm c}$ between pressure-induced superconductivity and Pt-doping-induced superconductivity.
For AuTe$_2$, $T_{\rm c}$ = 2.3~K at 2.34~GPa, which is the highest $T_{\rm c}$ under pressure, while $T_{\rm c}$ = 4.0~K for (Au$_{0.65}$Pt$_{0.35}$)Te$_2$.
One of the differences between the two systems is in the carrier number.
Pt doping at the Au site corresponds to hole doping.
Thus, $E_{\rm F}$ is shifted to the low-energy side by Pt doping and $D(E_{\rm F})$ increases, as shown in Fig.~\ref{Fig.5}.
Actually, the $\gamma_{\rm cal}$ in (Au$_{0.65}$Pt$_{0.35}$)Te$_2$ is larger than that in the high-pressure phase, as shown in Table~\ref{table:1}.
These results indicate a close relationship between $T_{\rm c}$ and $D(E_{\rm F}) \propto \gamma$.
Therefore, we try to calculate the $\gamma$ dependence of $T_{\rm c}$ in this system.
According to McMillan's formula\cite{W.L.McMillan_PR_1968}, $T_{\rm c}$ is determined as
\begin{align}
T_{\rm c} = \frac{\Theta_{\rm D}}{1.45}\exp\left[\frac{-1.04(1+\lambda)}{\lambda-\mu^{*}(1+0.62\lambda)}\right], 
\end{align}
where $\Theta_{\rm D}$ is the Debye temperature, $\lambda$ is the electron-phonon coupling constant, and $\mu^{*}$ is the Coulomb pseudopotential.
Note that both $\lambda$ and $\mu^{*}$ are related to $\gamma$.
We assume that $\lambda$ and $\mu^{*}$ are proportional to $\gamma$, and $\lambda$ and $\mu^{*}$ are represented by the normalized value $\frac{\gamma}{\gamma_{\rm Pt}}$, where $\gamma_{\rm Pt}$ is the $\gamma_{\rm cal}$ in (Au$_{0.65}$Pt$_{0.35}$)Te$_2$. 
[$\lambda = \lambda_0 \frac{\gamma}{\gamma_{\rm Pt}}$, and $\mu^{*} = \mu^{*}_0\frac{\gamma}{\gamma_{\rm Pt}}$, and $\lambda_0$ and $\mu^{*}_0$ do not change with pressure or Pt doping.]
In addition, Debye frequency does not change significantly with Pt doping, according to the specific heat measurements\cite{K.Kudo_JPSJ_2013}.
Then, we assume that $\Theta_{\rm D}$ does not change with pressure or Pt doping.
For (Au$_{0.65}$Pt$_{0.35}$)Te$_2$, we obtain $\lambda_{0}$ = 0.69 with the experimental values $T_{\rm c}$ = 4.0~K and $\Theta_{\rm D} = 187$~K\cite{K.Kudo_JPSJ_2013}, and the typical value $\mu^{*}_0$ = 0.13.
The calculated $T_{\rm c}$ based on the McMillan's formula can reproduce the experimental $T_{\rm c}$ in this system as shown in Fig.~\ref{Fig.6}.
In AuTe$_2$ at ambient pressure, $T_{\rm c}$ becomes almost zero owing to the small $D(E_{\rm F})$.
The variation in the electronic structure due to incommensurate modulation prevents the appearance of superconductivity since the calculated $T_{\rm c}$ becomes sufficiently large with $\gamma_{\rm cal}$ in the average structure of the low-pressure monoclinic phase.
Moreover, the maximum $T_{\rm c}$ in this system will be realized in (Au$_{0.65}$Pt$_{0.35}$)Te$_2$, since the Fermi energy in (Au$_{0.65}$Pt$_{0.35}$)Te$_2$ locates around the peak of the density of states, as shown in Fig.~\ref{Fig.5} (c).

In summary, we measured the pressure dependences of resistivity and ac susceptibility in AuTe$_2$.
In the low-pressure incommensurately modulated monoclinic phase, resistivity is strongly suppressed by applying pressure.
The $\gamma_{\rm cal}$ in the low-pressure monoclinic phase is clearly larger than that estimated from the specific heat measurement.
This disagreement suggests that the calculation of the average structure does not reproduce the actual band structure, implying the importance of Te$_2$ dimers due to the incommensurate modulation of the crystal structure for the electronic state.
In the high-pressure phase, which seems to correspond to the trigonal structure, zero resistivity and a diamagnetic shielding signal are observed at low temperatures, which clearly indicates the appearance of superconductivity.
For AuTe$_2$, $T_{\rm c}$ = 2.3~K at 2.34 GPa, which is the highest $T_{\rm c}$ under pressure, while $T_{\rm c}$ = 4.0~K for Pt-doped (Au$_{0.65}$Pt$_{0.35}$)Te$_2$.
The difference in $T_{\rm c}$ between the two systems can be explained by the difference in $D(E_{\rm F})$ based on McMillan's formula.


\section*{Acknowledgments}
This work was partially supported by the Kobe University LTM center,  a Grant-in-Aid for Scientific Research from the Japan Society for the Promotion of Science (JSPS), KAKENHI (B) (No. 24340085), a Grant-in-Aid for Challenging Exploratory Research (No. 246541065), and the Funding Program for World-Leading Innovative R\&D on Science and Technology (FIRST Program) from JSPS.
One of the authors (SK) is financially supported by a JSPS Research Fellowship.


\end{document}